# HISTEX (HISTory EXerciser) : A tool for testing the implementation of Isolation Levels of Relational Database Management Systems

Dimitrios Liarokapis
TEI of Epirus
Greece
dili@teiep.gr

Elizabeth O'Neil
University of Massachusetts Boston
USA
eoneil@cs.umb.edu

Patrick O'Neil
University of Massachusetts Boston
USA
poneil@cs.umb.edu

## ABSTRACT

We present a multi-process application called HISTEX (HISTory EXerciser), which executes input histories in a generic transactional notation on commercial DBMS platforms. HISTEX could be used to discover potential errors in the implementation of Isolation Levels by Relational Database Management Systems or cases where a system behaves over restrictively. It can also be used for performance measurements related to database workloads executing on real database systems instead of simulated environments. HISTEX has been implemented in C by utilizing Embedded SQL. However, many of its ideas could be reincarnated in new implementations that could rely on other database connectivity paradigms such as JDBC, JPA etc. We expect that by presenting some of the ideas behind its development we could re-invigorate some fresh interest and involvement in the research community regarding such tools.

## CSS CONCEPTS

Information Systems, Database Management Systems

## 1 INTRODUCTION

Isolation Levels have been introduced in Relational Database Management Systems (RDBMSs) in order to increase performance when absolute concurrency correctness is not necessary or when correctness can be guaranteed at the application level. The ANSI SQL standard [5] has provided definitions for four isolation levels: READ UNCOMITTED, READ COMMITED, REPEATABLE READ and SERIALIZABLE

There has been critique in the literature, about the clarity and generality of these definitions [1-2]. This raises concerns about the quality of the implementation of isolation levels by database vendors since it becomes more probable that the implementation of concurrency control could be sometimes incorrect or over-restrictive. By incorrect we mean that a database system could allow executions that should be proscribed by a given isolation level, leading to an unintentional corruption of the database or to the return of incorrect information. By over-restrictive we mean that the database system would not allow executions that are not proscribed by the isolation level at use. This would generally lead to reduced performance.

HISTEX was developed in order to assist coping with questions similar to: "Does a given database system implement Isolation Levels correctly?" or "Can we design a tool and methodologies to test the support of isolation levels by database management systems?"

One of the motivating ideas for creating HISTEX [9] was to develop a system that executes database concurrent scenarios similar to the ones used in the literature to argue about what transactional history should be permitted by a database scheduler.

HISTEX was developed for an NSF supported project [10] and has been partially documented at [6]. A Doctoral Poster was presented at VLDB 2001 [7] and it has been utilized for performance measurements for the work in [4].

In this paper, we highlight the HISTEX notation and its design and implementation and we explain how we have used it as a means to test and understand the implementation of isolation levels in commercial database management products.

## 2 HISTEX NOTATION

HISTEX executes input histories written in a generic transactional notation on commercial DBMS platforms. The HISTEX notation can be used by a researcher in much the same way as the classical transactional notation found in [3] to write down sequences of operations that model concurrent histories. HISTEX provides additional operations including simple ones such as Inserts, Deletes, and indivisible Read-Write Update operations ($RW_i(A)$), and more complex ones such as predicate evaluation, $PR(P...)$, which represent an Open Cursor operation and/or a sequence of Fetch operations from the Cursor.

While our HISTEX notation extends the number of operations from classical notation, it is nevertheless *generic*, meaning that it leaves details of operations as undefined as possible. By avoiding SQL-level specification, it allows researchers to concentrate on expressing a history, rather than becoming fixated on unimportant details.

The HISTEX program module interprets generic parameters such as i, A, and X in $R_i(A, X)$, assigning an operation with subscript i to a particular transaction thread, performing a Read (Select) operation of a particular row it consistently assigns the name A, and reading the (default) column value(s) of A into a value-variable it associates with the name X, one of multiple memory variables maintained by HISTEX.

The HISTEX notation has an *output history* format to represent the results of an input history execution on a specific database platform (a specific DBMS and Isolation Level). In the HISTEX output history form, values will be provided for rows/columns read and written, and types of failures in history execution will be noted.

## 2.1 HISTEX Data Model

By default, HISTEX interprets input histories in terms of a canonical relational table described as follows: T(reckey, recval,c2,c3,c4,c5,c6,c50,c100,k2,k3,k4,k5,k6,k50,k100)

The table T contains a parameterized number of rows (by default, the number is 200, but this can be altered to any multiple of 100). Columns k2 through k100 are indexed integer columns, where column kN has values 0, 1, 2, . . ., N-1 for successive rows, starting with the first row and extending through the last. Columns c2 through c100 have identical values with the corresponding k2 through k100 columns, but are not indexed (this will make a difference in predicate evaluation execution). The column named reckey is a primary key for the table, used to identify each individual row A, B, . . . used in an operation, and the column named recval is the default "value" of the row, which will be incremented by 1 when unspecified updates are performed. The values for reckey will be successively assigned to rows in T with values 100, 200, . . ., and recval values will be successively assigned with values 10000, 20000, . . ..

**Table 1: HISTEX Default Table**

| reckey | recval | c2 | c3 | … | c100 | k2 | k3 | … | k100 |
|---|---|---|---|---|---|---|---|---|---|
| 100 | 10000 | 0 | 0 |  | 0 | 0 | 0 |  | 0 |
| 200 | 20000 | 1 | 1 |  | 1 | 1 | 1 |  | 1 |
| 300 | 30000 | 0 | 2 |  | 2 | 0 | 2 |  | 2 |
| 400 | 40000 | 1 | 0 |  | 3 | 1 | 0 |  | 3 |
| 500 | 50000 | 0 | 1 |  | 4 | 0 | 1 |  | 4 |
| 600 | 60000 | 1 | 2 |  | 5 | 1 | 2 |  | 5 |
| . . . | . . . | .. | .. |  | . . . | .. | .. |  | . . . |

## 2.2 HISTEX Operations

Histex executes input histories that are a sequence of the operations described in this section. The declarative operations are listed first. These operations will not cause any database access. They are used to initialize variables that can be referenced in subsequent operations. The operations are briefly explained by examples. More detailed descriptions are included in [6].

**Predicate Declaration**. This operation will associate a predicate variable P with a predicate expression suitable for a SQL Where Clause. Example: PRED (P, "k2=1 and k3<2").

**Row Declaration.** This operation associates a row variable with a specific row. The row id is expected to be a value of the reckey column of the underlying table. Example: MAP(A, 100)

**Isolation Level Set Operation**. This operation is used to set the transaction isolation level for a particular transaction. It must be the first operation of a transaction. Currently the isolation levels that can be specified by this command depend on the underlying DBMS executing the history. Example: $IL_1$ (SR) will set the isolation level of transaction 1 to SERIALIZABLE (SR).

**Write Operation.** This operation models a blind update of a row. For example, when HISTEX interprets the operation $W_1$(A, 1001) it will identify the row variable A with a row (identified internally by its reckey). If A has already been identified with a row at some prior point in the history (even in an operation of a different transaction), then A will continue to be identified with the same row. To execute the operation W1(A, 1001), HISTEX is going to execute an SQL statement of the form: update T set recval = 1001 where reckey = 100.

**Read Operation.** This operation performs a read of a data item (row). For example $R_1$(A, X) will associate the row-variable A with a specific row in the underlying table in the same way as the Write operation described above. It will then open an SQL cursor for the prepared statement: select <column_name> into :value from T where reckey=100. The value retrieved by this operation will be assigned to the variable X.

**Insert Operation.** This operation will insert a new data item row in the underlying table. $I_1$(A) will associate the row-variable *A* to a new reckey value. Then it will insert a new row in the underlying table, which will contain the new reckey value and some default values for the rest of the columns. $I_2$(B;recval;k2;k3, 3000;0;2) will associate variable B to a new reckey value. It will then execute the SQL statement: insert into T (reckey, recval, k2, k3) values (350,3000,0, 2), in order to set specific values in the desired columns.

**Delete Operation.** This operation will delete a data item (row) from the underlying table.

**Predicate Read Operation.** The purpose for introducing this operation is to provide a way of producing predicate-read/item-write conflicts. $PR_1$(P;recval;1;A, X) will attempt to read one (1) row that matches the predicate P. In the current implementation, this operation will cause the opening of a cursor for selecting the recval column of the rows that match the predicate P. When the operation includes just a column name, as in this example, the reckey column is retrieved as well. The value of the reckey column will be registered in the variable A. The value of the recval will be registered in the variable X. $PR_1$(P;reckey;all) is going to retrieve all rows that satisfy the predicate P by reading the reckey value. Note that this is the most economical operation for accessing a predicate. If an index has been created for the reckey column, the operation can execute without having to retrieve the rows that satisfy the predicate; it only has to access the index. $PR_1$(P;count(*);1) will cause the execution of a SQL statement of the form: *select count(*) from T where P.* The database will return the number of rows that satisfy the predicate.

**Commit Operation.** $C_1$ will issue a COMMIT statement in the thread associated with transaction 1.

**Abort Operation.** $A_1$ is going to issue a ROLLBACK statement in the thread associated with transaction 1.

**Macro expanded SQL statements.** There are two operations: **EXECSQLI** and **EXECSQLS** that can be used to accommodate special cases.

For example in the history: PRED(P, k2=0 and k1=1), EXECSQLI$_1$(update T set recval = recval + 1 where %P), HISTEX is going to substitute %P with "k2=0 and k1=1" and then execute the derived statement as an operation of transaction 1. This operation can be used for executing what we call a set update operation – an operation that will update all the rows that satisfy a predicate.

For the input history: PRED(P, k2=0 and k1=1), EXECSQLS$_1$(select sum(recval) from T group by k1, k2 having %P), HISTEX is going to substitute %P with the corresponding predicate, and then execute the select statement by opening a cursor and fetching all the selected rows. Currently the operation does not return any values. It can be used to test whether SQL select statements that have not been mapped to other HISTEX operations acquire the necessary locks.

## 3 HISTEX IMPLEMENTATION

HISTEX is a multi-process application. It consists of two major modules: **monitor** and **thread**. At runtime there is a process executing the monitor module, and there can be several processes, each one executing a thread module.

The monitor process is the one that creates the thread processes. It is responsible for scanning and loading the input history, maintaining structures containing the state of the variables that are used in the history, and producing the output.

The thread module is responsible for interacting with a database system and provides the embedded SQL implementation of the HISTEX operations.

There is a communication protocol between the monitor and the threads. The monitor sends messages terminated with a newline character and the thread responds with SUCCESS or FAILURE. The thread's response also contains any value requested by the monitor, or some error message.

The current implementation of HISTEX is done for the UNIX operating system. In what follows, we describe some of the more complex parts of the implementation.

### 3.1 Communication between the monitor and the thread processes

We have developed a rather generic mechanism for implementing a system of a monitor process communicating with a group of thread processes. The system is developed in such a way that it could be utilized by other applications that follow a similar pattern.

In order to create a new thread the monitor uses the following function: Thread create_thread (void *call_thread(), void *parameters)

The first argument is a pointer to the function that will be called by the newly created thread. In the current application this function is the one that contains the Embedded SQL implementation of the HISTEX operations. The second argument is a pointer to a generic structure that can hold the parameters that the monitor would pass to the thread. Currently the only parameter passed is the default isolation level. The function will return the thread id encapsulated in the type Thread. Currently this type is implemented as an integer.

The communication between the monitor and the thread processes is implemented with UNIX pipes. When a new thread process is created, a pair of pipes is created between the monitor and the thread process, one for each direction of data flow.

A message can be sent to a thread by using the following function: Boolean send_to_thread (Thread threadId, char *buff). ThreadId is the thread to receive the message. The parameter buff points to the message which must be a newline terminated string.

The function: Boolean wait_for_thread(Thread threadId, int timeout, Boolean *toflag) is used by the monitor process in order to wait for a particular thread to respond. It will return FALSE if a system error occurred while waiting, and TRUE otherwise. The parameter threadId specifies the thread. The parameter timeout specifies a maximum number of seconds that the function will wait for the thread to respond. The parameter toflag will be set to TRUE if a timeout occurred while waiting, and to FALSE otherwise.

The function: Boolean wait_for_any_thread(Thread *threadId, int timeout, Boolean *toflag) is used to wait for any thread to respond. This is used when HISTEX is executed in the asynchronous mode. Under this mode it is possible that after HISTEX has submitted a history operation to an execution thread, it could proceed in processing the next operation even though the earlier one has not completed yet. After HISTEX has submitted all the possible operations, it calls this function to wait for a thread to send a response. The argument threadId will identify the thread.

For receiving a response from a thread the following function is used: Boolean receive_from_thread (Thread threadId, char *buff)

Finally the function: void finalize_threads () is used to terminate all thread processes.

### 3.2 Implementation of the Predicate Read Operation

We have implemented the predicate read (PR) HISTEX operation by using SQL cursors. Another approach could have been to just use a SELECT statement. Choosing a cursor implementation allows a wider range of testing scenarios such as the partial evaluation of a predicate. This is important in order to observe how a database system that implements Key-Value Locking behaves (e.g. we expect the locking to incrementally advance across a predicate set).

In such cases, it is possible that at some point in time a transaction $T_1$ has already opened a cursor and fetched some of the rows. At that moment, the database system receives an update operation from transaction $T_2$, which changes the matches of the predicate used by $T_1$. In a system that implements KVL locking [8] it is not certain that this operation will be blocked. It is possible that $T_1$ has already locked a range of values that contain some of the column values of the row updated by transaction $T_2$.

In this case the update of $T_2$ will be blocked. Another possibility is that transaction $T_1$ has not yet acquired any lock that could conflict with the update done by $T_2$. In that case $T_1$ could perform the update, and the scheduler will consider $T_1$ serialized before $T_2$. Later on when the scanning of the predicate operation reaches the range that could conflict with the update, it is possible that the operation will be blocked (in case $T_1$ has not committed yet).

The PR operation of HISTEX provides the ability to form such scenarios. Different instances of this command can be used by the same transaction in order to access a predicate in several steps. Each instance needs to reference the same predicate variable and can specify the number of rows that will be accessed each time. In order to implement this feature we are associating a predicate variable with some SQL cursor. The number of cursors that can be opened simultaneously is fixed. By convention, we use integer numbers to identify each cursor.

The first time a predicate variable is used by a PR operation, a cursor has not been opened yet. The message sent by the monitor process contains the value 0 as the required cursor id. When the thread processes the request, it will identify the next available cursor ID and will attempt to open a cursor.

To open cursors we use a switch statement where the case labels are the supported cursor ids. The following figure shows the C code for handling the opening of cursors:

```
switch(cursor_id) {

  case 1:
    EXEC SQL PREPARE S1 FROM :sql_stmt;
    EXEC SQL DECLARE C1 CURSOR FOR S1;
    EXEC SQL OPEN C1;
    break;
  case 2:
    EXEC SQL PREPARE S2 FROM :sql_stmt;
    EXEC SQL DECLARE C2 CURSOR FOR S2;
    EXEC SQL OPEN C2;
    break;

    ...
}
```

The code associates a cursor id with variables Sx and Cx for the corresponding prepared statement and cursor. The reason we use this technique is that it appears that we could not store cursor identifiers into an array and thus be able to dynamically access them

The cursor id of a recently opened cursor will be included in the thread's response. The monitor process will map the predicate variable used in the PR operation to this cursor id.

When the monitor encounters a subsequent PR operation referencing the same Predicate variable, it will extract the cursor id that was already mapped to this variable, and it will include it in the message sent to the thread. The following is the logic executed by the thread when a specific cursor id is provided:

```
switch (cursor_id) {

  /* arg3 indicates the rows to scan */
  /* aggrfl is set when an aggregate */
  /* operation is processed */

  case 1:

    if (!strcmp(arg3, "all")) {
      EXEC SQL WHENEVER NOT FOUND GOTO label_1_1;
      I = 0; for(;;) {
        if (aggrfl) {
          EXEC SQL FETCH C1 INTO :value;
        } else {
          EXEC SQL FETCH C1 INTO :key, :value;
        }
        i++;
    }
    label_1_1 :
    EXEC SQL CLOSE C1;
    free_cursorid(cursor_id);
    cursor_id = -1;

    } else {
      EXEC SQL WHENEVER NOT FOUND GOTO label_1_2;
      j = atoi(arg3);
      i = 0; while (i<j) {
        if (aggrfl) {
          EXEC SQL FETCH C1 INTO :value;
        } else {
          EXEC SQL FETCH C1 INTO :key, :value;
        }
        i++;
      }
      label_1_2:
      ;
    }
    break;

  case 2:
  /* same as in case 1
      (substitute C2 for C1 etc */
  …
```

In the preceding code, the different cases of the switch statement are identical except for the use of different names for some variables. If the PR operation requests that **all** rows are read, then the logic in the first conditional block will be executed and the whole rowset will be scanned. Having completed this, the cursor will be closed and the corresponding cursor id will be freed, becoming available for use by a different PR statement. Otherwise, only the specified number of rows will be retrieved.

Note that in the case that a specified number of rows is requested, the cursor will not be recycled even after all rows have been fetched. This behavior has been chosen so that it can be easily determined when a cursor has been closed.

## 3.3 Synchronous vs asynchronous execution mode

By default, HISTEX executes histories in a synchronous (serial) mode (i.e. an operation is processed after any preceding operation has been executed). This mode is the one used for the tests we are presenting later on. HISTEX also provides an asynchronous (concurrent) mode of execution, where operations can be submitted simultaneously to different threads. This is necessary when it is desirable to create concurrent workloads in a database. We have used this feature for implementing the performance measuring experiments in [4]. The current level of concurrency is that of one outstanding operation per transaction (i.e. the monitor submits any operation it encounters to the corresponding thread until it reaches an operation of a transaction with an unfinished operation). The asynchronous mode is enabled by using the HISTEX option (-c).

## 4 EXPERIMENTATION

We developed HISTEX with the intent of testing whether database vendors correctly implement isolation levels. A general approach that occurred to us was to create a large number of random input histories and execute them under different database systems and the isolation levels provided. The output histories could then be analyzed to determine whether those histories should have been produced by a given isolation level.

An approach to generate stochastic tests to determine when different database systems provided different answers to identical SQL statements has already been successful [11].

We considered using the characterization that appeared in [1] to determine when output histories were legal under various isolation levels. This work provides implementation-independent definitions of the isolation levels so that ORACLE multi-version concurrency can be treated similarly to the way locking concurrency is treated for other database systems. The paper uses a multi-version notation for defining histories and it defines the isolation levels based on the type of cycles that could be allowed in the output history and a few additional constraints.

However, there were several reasons that suggested we could not readily rely on the approach used in [1]. The main reason was that histories appearing in the paper also needed to specify the versions of the items that a system had chosen for every operation. Clearly the operations in our input histories could not specify particular versions, and a multi-version system, like ORACLE, did not report what versions were used. It could only report the values of the data items and the analyzer of the output history would have to determine the version.

The approach in [1] also relied on the existence of a version order of the data items included in the history, information that was not available in an output history from ORACLE, and there was no general way of deriving it. In addition, to determine the predicate dependencies, we would require knowledge of the whole database state at the moment the predicate operation occurred. In order to avoid performing a total read of the relations involved in the predicate, which is an action that would alter the meaning of the examined history, we would need to implement a complex mechanism for mirroring the database so that the versions of the rows in a table could be identified by querying this mirror.

Finally, in order to detect a predicate conflict, when examining the output history we should be able to reproduce the state of every row updated before and after the update takes place. Even though the update might specify only the value of a column being modified, in order to determine if the row update affects some predicate, the values of the rest of the columns must be available.

While it will be of research interest to consider ways of addressing the issues of dealing with multi-version concurrency, we decided to concentrate on testing isolation levels acting under locking concurrency. When we created HISTEX, all major commercial databases other than ORACLE were using locking, including DB2, Informix, and Microsoft's SQL Server.

For reasons that will become apparent in the sequel, we have decided to define a methodology that utilizes assumptions of concurrency control mechanisms about the underlying database system. In this way we can considerably simplify the testing cases.

Instead of creating histories that would try to produce the phenomena (i.e. the output patterns) described in [2] or the cycles in [1], we determined that it would be sufficient to check whether the locking protocol is implemented according to Table 2.

Table 2 provides definitions of isolation levels based on the pairs of concurrent conflicting operations that should be avoided. These are actually the effects that a locking scheduler should have.

**Table 2: Isolation Levels defined in terms of prohibited concurrent pairs of conflicting operations**

| Locking Isolation Level | Concurrent pairs of conflicting operations that should be avoided |
|---|---|
| READ UNCOMMITTED | There are no conflicting operations. Transactions are READ ONLY |
| READ COMMITTED | $W_1(A)\ W_2(A)$<br>$W_1(A)\ R_2(A)$<br>$W_1(A\ \text{changes}\ P)\ PR_2(P)$ |
| REPEATABLE READ | All the above and:<br>$R_1(A)\ W_2(A)$ |
| SERIALIZABLE | All the above and:<br>$PR_1(A)\ W_2(A\ \text{changes}\ P)$ |

A W operation in this table stands for any operation that performs a Write (i.e. W, RW, I, D). We will rely on this table to define a plan for testing the correctness of isolation levels by database systems. [6] provides a proof of a theorem stating that: *If a database system prevents the concurrent operations in Table 2, then it implements the isolation levels correctly. By "correctly", we mean that at a given isolation level there will be no phenomenon occurring that is proscribed by that level.*

### 4.1 Creating Testing Plans

For testing the READ UNCOMMITTED level we need to consider that this level should be used by READ ONLY transactions and there are no locks required [5]. We can use histories of the form: $R_1(A)R_1(B)C_1W_2(A)\ R_3(A, A0)\ W_3(B, A0)$ $C_3\ A_2\ R_4(A)\ R_4(B)\ C_4$ to check whether transaction 3 will be able to persist an update performed by the aborting transaction 2. Our experience on some commercial database products indicates that transactions that have started at the READ UNCOMMITTED level can perform updates.

For testing the remaining isolation levels READ COMMITTED (RC), REPEATABLE READ (RR) and SERIALIZABLE (SR), we have created a generator utility that can produce histories with conflicting pairs of operations. We can identify the classes of conflicting pairs listed in Table 3.

**Table 3: History Classes**

| History Class | Type of Conflict |
|---|---|
| w_w | item write / item write |
| w_r | item write / item read |
| r_w | item read / item write |
| w_pr | item write / predicate read |
| pr_w | predicate read / item write |

After executing the input histories, we can analyze the output histories to detect if the executions are according to the types of conflicting operations that are permitted as indicated in Table 4.

**Table 4: History Class Operations permitted under combinations of Isolation Level settings**

| History Class | Isolation Levels (RC or above) allowing execution |
|---|---|
| w_w | None |
| w_r | None |
| w_pr | None |
| r_w | RC_RC, RC_RR, RC_SR |
| pr_w | RC_RC, RC_RR, RC_SR, RR_RC, RR_RR, RR_SR |

We have implemented a template based generator to facilitate the creation of input histories containing all possible types of conflicting pairs of operations and we were able to execute them on several commercial database products and under all possible combinations of isolation levels and under the presence or absence of primary key constraints and indexes [6].

An interesting observation among our findings was that in a database product the isolation level that corresponds to the READ COMMITTED level in the ANSI SQL specification, would allow a transaction to observe an uncommitted state of the database.

This was happening in cases where a transaction T1 would force a row out of a predicate P, and before this transaction committed another transaction T2 accessing the rows in predicate P would not see this row.

It was also interesting to notice that some behavior that could be considered erroneous in a database system was fixed in a subsequent version.

In addition to incorrect behavior, our analysis also detected cases where the underlying database systems behaved over-restrictively.